\documentstyle[tighten,preprint,aps,epsf,prd]{revtex}
\begin{document}

\def\S{Schwarzschild\enspace}
\def\L{\Lambda}
\def\P{{\tilde P}_R}

\title{Microscopic Black Hole Pairs in Highly-Excited States}

\author{Jarmo M\"akel\"a\footnote{Electronic address: jarmo.makela@phys.jyu.fi} and Pasi Repo\footnote{Electronic address: pasi.repo@phys.jyu.fi}}
\address{Department of Physics, University of Jyv\"askyl\"a, P.O. Box 35, FIN-40351 Jyv\"askyl\"a, Finland}
\date{June 20, 2000}
\maketitle
\begin{abstract}
We consider the quantum mechanics of a system consisting of two identical,
Planck-size Schwarzschild black holes revolving around their common center of
mass. We find that even in a very highly-excited state such a system has very sharp, discrete energy eigenstates, and the system performs very rapid transitions from a one stationary state to another.
For instance, when the system is in the 100th excited state, the life times of the energy eigenstates are of the order of $10^{-30}$ s, and the energies of gravitons released in transitions between nearby states are of the order of $10^{22}$ eV.
\end{abstract}


\narrowtext

\def\R{{\hbox{R}}}

Some years ago Hawking and others introduced an interesting idea about a possible
spontaneous creation of virtual black hole pairs.~\cite{Haw1,Haw2,Haw3,Haw4,Haw5} This process would be
analogous to the spontaneous creation of electron-positron pairs in quantum
electrodynamics, and the members of the black hole pairs would presumably be
Planck-size objects. As it is well known, electrons and positrons may
sometimes, in very favourable conditions, form a system called positronium, in
which an electron and a positron revolve around their common center of mass.
The possibility of a spontaneous creation of black hole pairs, then, gives 
rise to some very interesting questions: Could the black hole pairs sometimes
form systems, analogous to positronium, where two microscopic black holes
revolve around each other? What are the possible quantum states of such
systems? What happens when the system performs a transition from one quantum
state to another? Could one observe these transitions? 

These are the questions which will be addressed in this paper. We shall
consider the simplest  possible case where the members of the pair are
Schwarzschild black holes, both having the same mass $M$. In general, the
problem of quantization of a system containing a black hole pair presents
immense difficulties: Indeed, even the classical solution to Einstein's
equations describing a spacetime containing two black holes is unknown.
However, if the two microscopic black holes are sufficiently far away from each
other, they can be considered, to a very good approximation, as point-like
particles moving with non-relativistic speeds. Moreover, the gravitational
interaction between the holes can be described by the good old Newtonian theory
of gravitation. Hence, the quantum mechanics of the black hole pair is described
by the non-relativistic Schr\"{o}dinger equation:
\begin{equation}
(-{{\hbar^2}\over{2\mu}}\nabla^2 - {{GM^2}\over r})\psi = E\psi \, ,
\label{Sch}
\end{equation}
where $\mu:={1\over 2}M$ is the reduced mass of the system, and $E$ is its total
energy. $r$ is the distance between the black holes.

Our first task is to check whether the approximations we just made
are valid. As it is well-known from elementary quantum mechanics, 
in stationary states the energy eigenvalues calculated from Eq.~(\ref{Sch}) are
\begin{equation}
E_n = -{1\over 2}{{GM^2}\over{\hbar^2 n^2}}\, ,
\label{energy}
\end{equation}
where $n=1,2,3,...$\,. In these states the expectation values of $r$ are
\begin{equation}
\langle r\rangle_{n,l} = {{2\hbar^2}\over{GM^3}}n^2\lbrace 1 + {1\over
2}\lbrack 1 - {{l(l+1)}\over{n^2}}\rbrack\rbrace \, ,
\label{expect}
\end{equation}
where $l=0,1,2,...,(n-1)$ is the angular momentum quantum number of the system.
Our approximations are valid if
\begin{equation}
{{2\hbar^2}\over{GM^3}}n^2 \gg R_S\, ,
\label{crit1}
\end{equation}
where
\begin{equation}
R_S := {{2GM}\over{c^2}}
\label{RS}
\end{equation}
is the Schwarzschild radius of a black hole with mass $M$. Comparing Eqs.~(\ref{crit1})
and~(\ref{RS}) we find that we must have
\begin{equation}
{1\over{\sqrt{n}}}M\ll M_{Pl}\, ,
\label{crit2}
\end{equation}
where
\begin{equation}
M_{Pl} := \sqrt{{{\hbar c}\over G}} \approx 2.2 \times 10^{-8}~ \mathrm{kg}
\label{Mpl}
\end{equation} 
is the Planck mass. Hence, either $n$ must be very big or $M$ must be much
smaller than the Planck mass. The average velocity on the orbit is
\begin{equation}
\langle v_{ave}\rangle_n = ({M\over{M_{Pl}}})^2{1\over n}c\, .
\label{velocity}
\end{equation}
As one can see, the black holes move with non-relativistic speeds if Eq.~(\ref{crit2})
holds. In what follows, we shall always assume that $M$ is smaller than, or
equal to, the Planck mass, and $n$ is very big. In other words, we shall
consider microscopic black hole pairs in highly-excited states.

Consider next the transitions between highly-excited stationary
states. It follows from Eq.~(\ref{energy}) that the energy released when the system
performs a transition where $n$ is reduced by one is
\begin{equation}
E_n - E_{n-1} \approx {{G^2M^5}\over{\hbar^2 n^3}} =
{1\over{n^3}}({M\over{M_{Pl}}})^4Mc^2\, .
\label{DeltaE}
\end{equation}
For instance, if $n$ is around ten and $M$ is the Planck mass, the energy
released is $10^{-3}$ times the Planck energy, or $10^6$ J, which is about the
same as the energy needed when an automobile is accelerated from rest to the
velocity of 100 km/h. As one can see, enormous energies could, in principle, be
stored in systems containing microscopic black hole pairs.

Since the holes are assumed to be uncharged, one may expect that the
main reason for transitions between stationary states is the quantum
fluctuation of the gravitational field between the holes: the quantum
fluctuation of the gravitational field perturbes the stationary states, a
transition occurs, and a graviton is emitted or absorbed spontaneously. Since
the black holes are assumed to be in a highly-excited state, and therefore
relatively far away from each other, we may use the linear field approximation
when investigating the perturbative effects caused by the quantum fluctuations
of the gravitational field.

The Lagrangian of a point particle moving in a weak gravitational
field $h_{\mu\nu}$ is
\begin{equation}
L=-M\sqrt{(\eta_{\mu\nu} +
h_{\mu\nu})\,{{dx^\mu}\over{dt}}\,{{dx^\nu}\over{dt}}}\, .
\label{lagrangian}
\end{equation}
Under the assumptions that $\vert h_{\mu\nu}\vert\ll 1$, and the particle moves very
slowly we may write Eq.~(\ref{lagrangian}), in SI units:
\begin{equation}
L \approx -Mc^2 + {1\over 2}M(\delta_{jk} - h_{jk}){\dot x}^j{\dot x}^k -
Mch_{0j}{\dot x}^j - {1\over 2}Mc^2h_{00}\, ,
\label{slowLag}
\end{equation}
where $j,k=1,2,3$ and ${\dot x}^j:=dx^j/dt$. Dropping the term $-Mc^2$, which
is a mere constant, we may infer that, in the center of mass coordinates, the
Lagrangian of the black hole pair can be written, in effect, as
\begin{eqnarray}
L &=&{1\over 2}\mu\lbrack \delta_{jk} - 
{1\over 2}h_{jk}({{\vec r}\over 2},t) -
{1\over 2}h_{jk}(-{{\vec r}\over 2},t)\rbrack{\dot x}^j{\dot x}^k \nonumber \\
           &-&\mu c\lbrack h_{0j}({{\vec r}\over 2},t) - h_{0j}(-{{\vec r}\over
2},t)\rbrack{\dot x}^j \nonumber \\
           &-&\mu c^2\lbrack h_{00}({{\vec r}\over 2},t) + h_{00}(-{{\vec
r}\over 2},t)\rbrack\, ,
\label{Leff}
\end{eqnarray}
where ${\vec r}$ is the vector joining the hole 1 to the hole 2, and $x^j$'s
are defined such that ${\vec r}=x^j{\hat e}_j$, where ${\hat e}_j$'s are orthonormal basis vectors. 

Now, it is easy to see that the terms proportional to $h_{00}$
represent the gravitational potential energy of the system. These terms should
give, when the system is in a highly-excited state, the Newtonian potential
energy between the holes. The terms proportional to $h_{0j}$ and $h_{jk}$, in
turn, are related to the quantum fluctuations of the gravitational field. If we
assume that transitions from a one stationary state to another are associated with
spontaneous emissions or absorptions of gravitons, the only remaining terms, in
addition to the Newtonian potential energy, are the terms proportional to
$h_{jk}$. That is because the "scalar" and the "vector" gravitons proportional to
$h_{00}$ and $h_{0j}$, respectively, can be gauged away, and the physical
gravitons correspond to the $(jk)$-components of the field $h_{\mu\nu}$.
Therefore, the Lagrangian of our system interacting with spontaneously emitted
or absorbed gravitons is
\begin{equation}
L = {1\over 2}\mu\lbrack\delta_{jk} - {1\over 2}h_{jk}({{\vec r}\over 2},t) -
{1\over 2}h_{jk}(-{{\vec r}\over 2},t)\rbrack{\dot x}^j{\dot x}^k +
{{GM^2}\over r}\, .
\label{Lint}
\end{equation}
Hence we find, under the assumption that $h_{jk}$ is small, that the Hamiltonian
operator of the black hole pair interacting with gravitational radiation is
\begin{equation}
{\hat H} = {{{\hat{\vec p}}^2}\over{2\mu}} - {{GM^2}\over{\hat r}} + {\hat
H}_{int}\, ,
\label{Hamilton}
\end{equation}
where
\begin{equation}
{\hat H}_{int} := {1\over{4\mu}}\lbrack {\hat h}^{jk}({{\vec r}\over 2},t) +
{\hat h}^{jk}(-{{\vec r}\over 2},t)\rbrack{\hat p}_j{\hat p}_k\, ,
\label{Hinter}
\end{equation}
and $p_j$ is the canonical momentum conjugate to $x^j$. In these equations,
objects equipped with hats are operators replacing the corresponding classical
quantities. 

The transition rate $\R_{i\rightarrow f}$ from the initial state $\vert i\rangle$ to
the final state $\vert f\rangle$ can be evaluated by means of the Golden Rule~\cite{Messiah}.
Evaluation of the transition rate involves quantization of the linearized
gravitational field in the radiation gauge where $h_{0\mu}=0$ for every
$\mu=0,1,2,3$. This has been performed in the Appendix~\ref{app:A}, and we find that if
exactly one graviton with angular frequency $\omega$ is emitted in a
transition, then
\begin{equation}
\R_{i\rightarrow f} = {{G\omega}\over{2\pi c^5\hbar\mu^2}}\vert\langle f\vert \cos({1\over
2}{\vec k}\cdot{\vec r})\epsilon^{ab}{\hat p}_a{\hat p}_b\vert
i\rangle\vert^2\,d\Omega\, ,
\label{transitrate'}
\end{equation}
where ${\vec k}$ is the wave vector of the graviton, $\epsilon^{ab}$ its
polarization tensor, and $d\Omega$ is the solid angle into which the graviton
emerges. The polarization tensors $\epsilon^{(1)ab}$ and $\epsilon^{(2)ab}$
corresponding to the two physical polarizations of the graviton have been
chosen such that
\begin{equation}
\epsilon^{(\lambda')ab}\epsilon^{(\lambda)}_{ab} =
2\delta^{\lambda'\lambda}
\label{polchoose}
\end{equation}
for every $\lambda', \lambda=1,2$. In the lowest order approximation we can
write:
\begin{equation}
\R_{i\rightarrow f} = {{G\omega}\over{2\pi c^5\hbar\mu^2}}\vert\langle
f\vert\epsilon^{ab}{\hat p}_a{\hat p}_b\vert i\rangle\vert^2\,d\Omega\, .
\label{transitrate}
\end{equation}
Transitions corresponding to this expression may be viewed as gravitational
analogues of the E1 transitions in atomic physics. We shall therefore call
them as G1 transitions.

       The next task is to integrate Eq.~(\ref{transitrate'}) over all the possible
directions into which the gravitons may emerge. This has been done in
Appendix~\ref{app:B}. If the initial state $\vert i\rangle$ and the final state
$\vert f\rangle$ are taken to be solutions to the Schr\"{o}dinger equation~(\ref{Sch})
of the black hole pair, and one takes into account the fact that gravitons have
exactly two independent polarizations, one finds that the integrated
transition rate from the state $\vert i\rangle$ to the state $\vert
f\rangle$ in G1 transitions is
\begin{equation}
R_{i\rightarrow f} = {{4G\omega}\over{3 c^5\hbar\mu^2}}\vert\langle f\vert{\hat
p}_1^2\vert i\rangle\vert^2\, .
\label{transrate}
\end{equation}

From this expression one can obtain the G1 selection rules for the black
hole pair. In other words, one finds the transitions for which the
integrated G1 transition rate~(\ref{transrate}) is non-zero. Since gravitons may be
viewed as spin two particles, one might expect that for allowed G1
transitions the angular momentum quantum number $l$ as well as the
corresponding azimuthal quantum number $m_l$ could change only by zero, or
by plus or minus two. A detailed investigation, which has been performed
in Appendix~\ref{app:C}, shows that this is indeed the case: the G1 selection rules
are:
\begin{mathletters}
\label{selrules}
\begin{eqnarray}
\Delta l &=& 0,\pm 2\, ,\\
\Delta m_l &=& 0,\pm 2\, .
\end{eqnarray}
\end{mathletters}
 
       It is straightforward, although very laborious, to calculate the
transition rates in allowed G1 transitions from Eq.~(\ref{transrate}). This arduous
task, with explicit expressions for transition rates in different G1
transitions, has been performed in Appendix D. For instance, if the
principal quantum number $n$ is of the order of 100, then it turns out
that the transition rates in transitions between nearby states are of the
order of $10^{28}$ 1/s, if $\Delta l=0,-2$, and of the order of
$10^{24}$ 1/s, if $\Delta l=+2$, provided that we assume that the mass $M$ is equal to
the Planck mass $M_{Pl}$. Transitions between nearby states, however, are
not necessarily the most favourable ones. In Figs. 1, 2, and 3 we have
plotted the transition rates as functions of $k$, the difference between the
initial and the final values of the principal quantum number $n$. One
finds, for instance, that when $\Delta l = \Delta m_l=-2$, and $l_i=m_{l_i}=50$, then the most favourable transitions are those where $k\approx 20$, and if, for the same values of $l_i$ and $m_{l_i}$, $\Delta l=0$ and $\Delta m_l=-2$, then for the most favourable transitions $k\approx 10$. If $\Delta l=+2$ and $\Delta m_l=0$, then $k=3$ for the most favourable transitions.

       As we have seen, Planck-size black hole pairs perform extremely
rapid transitions from one stationary state to another. However, the problem is,
whether the transitions are propably too rapid so that one cannot
meaningfully talk about discrete energy eigenstates at all. To answer this
question, one must calculate the life time of an initial state with big $n$,
when all the allowed G1 transitions are taken into account. The life time of
an initial state, in turn, is the inverse of the sum of the transition
rates of all the allowed G1 transitions. Using the formulas of Appendix D,
and assuming that transitions from higher to lower states are dominant, one
finds that the life time of an initial state with $n$ of the order of 100,
is of the order of
\begin{equation}
\tau_{100} \sim 10^{-30}~\mathrm{s}\, ,
\label{lifet}
\end{equation}
if $M=M_{Pl}$. This result settles, within the approximations made in this
paper, the question about the existence of sharp energy eigenstates.
According to Heisenberg's uncertainty principle the natural line width of
a state with life time $\tau_n$ is
\begin{equation}
\delta E_n \sim {\hbar\over{\tau_n}}\, ,
\label{linew}
\end{equation}
and therefore
\begin{equation}
\delta E_{100} \sim 10^{-4}~\mathrm{J}\, .
\label{linew2}
\end{equation}
However, the energy difference between nearby states is, according to
Eq.~(\ref{DeltaE}):
\begin{equation}
E_{100} - E_{99} \approx 2\times 10^3~\mathrm{J} \, ,
\label{Ediff}
\end{equation}
and therefore:
\begin{equation}
{{\delta E_{100}}\over{E_{100} - E_{99}}} \sim 10^{-7}\, .
\label{Ediff2}
\end{equation}
In other words, the uncertainty of the energy eigenvalues of the system is
very much smaller than the energy difference between nearby states. Hence,
the energy spectrum appears to be discrete, at least as far as one can
trust in the approximations made in this paper.  The transitions between
discrete energy eigenstates are extremely rapid, and the energies released
are enormous. 

       In this paper  we have investigated microscopic Schwarzschild black hole
pairs revolving around their common center of mass. Considering the holes as
point-like objects interacting with the Newtonian gravitational force, we
quantized the system and studied the stationary energy levels when the system
is in a highly-excited state. We then calculated, by means of perturbative
methods, the transition rates and life times in a certain class of transitions.
These calculations were based on the quantization of the linearized
gravitational field in the radiation gauge. We obtained, in the lowest
order approximation, explicit expressions for the transition rates and
evaluated the transition rates numerically when the pricipal quantum
number $n\sim 100$, and the mass of a microscopic black hole is assumed to
be one Planck mass. We found that the transition rates are of the order of
$10^{24-28}$ 1/s for allowed transitions, and the life times of energy
eigenstates are of the order of $10^{-30}$ s. Gravitons with energies of
the order of $10^{22}$ eV, or even greater, are emitted in these very rapid
transitions between discrete energy eigenstates. 

       No doubt, one might be justified to have some feelings of suspicion towards the
validity of the approximations on which this paper is based. Indeed, it may
well be somewhat daring to apply Newton's ancient theory of gravitation to
microscopic black holes! However, if one accepts the view that, in the
classical level, Einstein's general theory of relativity is the correct theory
of gravitation (so far we have no experimental evidence suggesting that the
things could be otherwise), having Newton's theory as its non-relativistic
limit, one is also forced to accept the "Newtonian approximation" made in
this paper: Holes with $M=M_{Pl}$ move, if $n\sim 100$, with velocities
which are of the order of $0.01c$ and the expectation value of their
mutual distance is about $10^4$ times their Schwarzschild radius. Of
course, one could calculate general relativistic corrections to energy
levels and transition rates but as far as one is interested in mere
order-of-magnitude estimates, Newton's theory should be 
sufficiently accurate. Quite another matter is, whether the first order
perturbative quantum theory of linearized general relativity is
sufficiently accurate an approximation of quantum gravity for the
evaluation of transition rates and life times. As far as the energies of
the emitted or absorbed gravitons are well below the Planck energy, which
is the case at least for transitions between nearby states when $n\sim
100$, however, one might be inclined to rely on the approximations made in this
paper. 
  
        One of the basic lessons one can learn from this paper is that
enormous energies could be relased by means of the quantum effects of the
gravitational field: The energies released in transitions between the energy
eigenstates of the microscopic black hole pair are, even in the transitions
between nearby states when $n\sim100$, about fourteen orders of magnitude
greater that the energies typically released in nuclear phenomena, and yet we are
talking about a microscopic system.

        It appears to us that the only conceivable way how microscopic
black hole pairs could be formed is a pairing of black hole remnants which are
left as the end products of Hawking evaporation of primordial black holes. The
existence of black hole remnants is a controversial question~\cite{Wald}, although it is
possible to construct mathematically consistent quantum theories of black holes
where the ground state energy of the hole is positive (see Refs.~\cite{Louko,Makela}). If such
remnants exist, however, it is possible  that some of them form pairs. Our
calculations show that if the masses of the remnants are of the order of
Planck mass, extremely energetic gravitons are emitted in
very rapid transitions from one state to another. Some of these very energetic
gravitons, in turn, may materialize into observable particles producing cosmic
rays with very high energies. Measuring the amount of very energetic
cosmic rays could therefore be used when trying to estimate the abundance of
black hole remnants in the universe. If there are black hole pairs formed by
black hole remnants with masses of the order of Planck mass, one expects to
observe cosmic rays with energies of the order of $10^{20}$ eV, or even
greater.

\acknowledgments

We are grateful to Matias Aunola, Markku Lehto and Jorma Louko for usefull discussions and constructive criticism during the preparation of this paper. 

\appendix
\section{Quantization of Linearized Gravity in the Radiation Gauge}
\label{app:A}

The Lagrangian density of the linearized gravitational field $h_{\mu\nu}$
can be written, when the Hilbert gauge condition
\begin{equation}
\partial_\mu(h^{\mu\nu} - {1\over 2}\eta^{\mu\nu}h) = 0
\label{Hilbgau}
\end{equation}
is used, as $(c=1)$~\cite{Feynman}:
\begin{equation}
{\cal L} = {1\over{64\pi G}}(\partial_\lambda
h_{\mu\nu}\partial^\lambda h^{\mu\nu} - 2\partial^\mu
h_{\mu\nu}\partial_\lambda h^{\nu\lambda} + 2\partial^\mu h_{\mu\nu}\partial^\nu
h - \partial_\nu h\partial^\nu h) \, .
\label{Lagdens1}
\end{equation}
In the radiation gauge we have, in addition
\begin{equation}
h^{\mu0} = 0\,\,\, \forall\mu=0,1,2,3\, ,
\label{radgau}
\end{equation}
and the Lagrangian density can be written in the form:
\begin{equation}
{\cal L} ={1\over{64\pi G}}(\dot{h}_{mn}\dot{h}^{mn} + \partial_l
h_{mn}\partial^l h^{mn} - \partial_n h\partial^n h)\, ,
\label{Lagdens2}
\end{equation}
where $\dot{h}_{mn}:=\partial h_{mn}/\partial t$, and the Latin indecies take
the values 1,2,3. The canonical momentum conjugate to $h_{ab}$ is therefore
\begin{equation}
p^{ab} := {{\partial\cal{L}}\over{\partial\dot{h}_{ab}}} = {1\over{32\pi
G}}\dot{h}^{ab}\, .
\label{conjmom}
\end{equation}

    As the first step towards quantization we write the field $h_{ab}$ as a
Fourier expansion:
\begin{equation}
h_{ab}(t,\vec{r}) = \sum_{\vec{k}}\sum_{\lambda=1}^2 \epsilon_{ab}^{(\lambda)}
(u_{\vec{k}}(t,\vec{r})a_{\vec{k}}^{(\lambda)} +
u_{\vec{k}}^{*}(t,\vec{r})a_{\vec{k}}^{(\lambda)\dagger})\, .
\label{Fourierexp}
\end{equation}
In this equation, the sum is taken over the wave vectors $\vec{k}$ and the
polarizations $\lambda$. $\epsilon_{ab}^{(\lambda)}$ is the polarization
tensor, $a_{\vec{k}}^{(\lambda)}$ and $a_{\vec{k}}^{(\lambda)\dagger}$ are the
Fourier coefficients, and the functions $u_{\vec{k}}$ are orthonormal wave
modes:
\begin{equation}
u_{\vec{k}}(t,\vec{r}) = N_{\vec{k}} e^{i(\vec{k}\cdot \vec{r} - \omega
t)}\, ,
\label{wavemodes}
\end{equation}
where $N_{\vec{k}}$ is a normalization constant, and $\omega$ is the angular
frequency of the graviton. 
      
              Introducing periodic boundary conditions in a box with edge $L$:
\begin{mathletters}
\label{boucond}
\begin{eqnarray}
h_{ab}(0,y,z,t) &=& h_{ab}(L,y,z,t)\, , \\
           h_{ab}(x,0,z,t) &=& h_{ab}(x,L,z,t)\, ,\\
           h_{ab}(x,y,0,t) &=& h_{ab}(x,y,L,t)\, , 
\end{eqnarray}
\end{mathletters}
we find that the possible values of $k_x$, $k_y$ and $k_z$ are
\begin{mathletters}
\label{quantnumb}
\begin{eqnarray}
           k_x &=& n_x{{2\pi}\over L}\, ,\\
           k_y &=& n_y{{2\pi}\over L}\, ,\\
           k_z &=& n_z{{2\pi}\over L}\, ,
\end{eqnarray}
\end{mathletters}
where $n_x$, $n_y$ and  $n_z$ are integers. Introducing, moreover, the inner
product between wave modes,
\begin{equation}
\langle u_{\vec{k}'}\vert u_{\vec{k}}\rangle :=
-i\int_0^Ldx\int_0^Ldy\int_0^Ldz\,(u_{\vec{k}'}\dot{u}_{\vec{k}}^* -
\dot{u}_{\vec{k}'}u_{\vec{k}}^*)\, ,
\label{innerprod}
\end{equation}
together with the requirement that the wave modes are orthonormal:
\begin{equation}
\langle u_{\vec{k}'}\vert u_{\vec{k}}\rangle =
\delta_{\vec{k}'\vec{k}}\, ,
\label{orthonormality}
\end{equation}
implies
\begin{equation}
u_{\vec{k}}(t,\vec{r}) = (2L^3\omega)^{-1/2}e^{i(\vec{k}\cdot\vec{r} - \omega
t)}\, .
\label{normalmode}
\end{equation}

        Consider now the polarizations of gravitons. The Hilbert gauge
condition~(\ref{Hilbgau}), together with the radiation gauge condition~(\ref{radgau}) implies
\begin{equation}
\epsilon_{ab}^{(\lambda)}k^b = 0\, ,
\label{innerprod2}
\end{equation}
for all $a=1,2,3$ and $\lambda=1,2$. To satisfy this condition, we may choose
the polarization tensors $\epsilon_{ab}^{(\lambda)}$ such that
\begin{equation}
\epsilon_{ab}^{(\lambda)}\epsilon^{(\lambda')ab} =
2\delta^{\lambda\lambda'}\, ,
\label{polcond}
\end{equation}
for all $\lambda,\lambda'=1,2$. For instance, we may choose:
\begin{mathletters}
\label{poltensors}
\begin{eqnarray}
\left(\epsilon_{ab}^{(1)}\right) &=& \left(\begin{array}{ccc}
			1&0&0 \\
                                                0&-1&0 \\
                                                0&0&0 
		\end{array}\right)\, ,\\
\left(\epsilon_{ab}^{(2)}\right) &=&  \left(\begin{array}{ccc} 
			0&1&0  \\
                                                1&0&0 \\
                                                0&0&0 
		\end{array}\right)\, .
\end{eqnarray}
\end{mathletters}
In that case the gravitons propagate to $z$-direction. Using Eqs.~(\ref{orthonormality}) and~(\ref{polcond})
we can write the Fourier coefficients $a_{\vec{k}}^{(\lambda)}$ and 
$a_{\vec{k}}^{(\lambda)\dagger}$ as:
\begin{mathletters}
\label{Fouriercoeff}
\begin{eqnarray}
a_{\vec{k}}^{(\lambda)\dagger} &=& -{i\over
2}\epsilon^{(\lambda)ab}\int_0^L dx\int_0^L dy \int_0^L
dz\,(u_{\vec{k}}\dot{h}_{ab} - \dot{u}_{\vec{k}}h_{ab})\, ,\\
\label{adagger}
           a_{\vec{k}}^{(\lambda)} &=& {i\over
2}\epsilon^{(\lambda)ab}\int_0^L dx\int_0^L dy \int_0^L
dz\,(u_{\vec{k}}^*\dot{h}_{ab} - \dot{u}_{\vec{k}}^*h_{ab})\, .
\label{a}
\end{eqnarray}
\end{mathletters}

We now proceed to quantization. It follows from Eq.(~\ref{conjmom}) that the canonical
equal time commutation relations between the field operators $\hat{h}_{ab}$ 
and their conjugates $\hat{p}^{ab}$,
\begin{mathletters}
\label{ETCCR}
\begin{eqnarray}
\lbrack \hat{h}_{ab}(t,\vec{r}'),\hat{p}^{cd}(t,\vec{r})\rbrack & =&
i\hbar\delta^3(\vec{r}',\vec{r})\delta^c_a\delta^d_b\, , \\
\label{hp}
          \lbrack \hat{h}_{ab}(t,\vec{r}'), \hat{h}_{cd}(t,\vec{r})\rbrack &=&
\lbrack \hat{p}^{ab}(t,\vec{r}'),\hat{p}^{cd}(t,\vec{r})\rbrack =
0\, , 
\label{hh}
\end{eqnarray}
\end{mathletters}
can be written as:
\begin{mathletters}
\label{ETCCR2}
\begin{eqnarray}
\lbrack \hat{h}_{ab}(t,\vec{r}'),\dot{\hat{h}}^{cd}(t,\vec{r})\rbrack & =&
i32\pi G\hbar\delta^3(\vec{r}',\vec{r})\delta^c_a\delta^d_b\, ,\\
\label{hhdot}
          \lbrack \hat{h}_{ab}(t,\vec{r}'), \hat{h}_{cd}(t,\vec{r})\rbrack &=&
\lbrack \dot{\hat{h}}^{ab}(t,\vec{r}'),\dot{\hat{h}}^{cd}(t,\vec{r})\rbrack =
0\, .
\label{hh2}
\end{eqnarray}
\end{mathletters}
Using Eqs.~(\ref{polcond}) and ~(\ref{Fouriercoeff}) we find that the commutation relations between the
operators $\hat{a}_{\vec{k}}^{(\lambda)\dagger}$ and $\hat{a}_{\vec{k}}^{(\lambda)}$ are:
\begin{mathletters}
\label{comrel}
\begin{eqnarray}
\lbrack \hat{a}_{\vec{k}'}^{(\lambda')},
\hat{a}_{\vec{k}}^{(\lambda)\dagger}\rbrack &=& 16\pi
G\hbar\delta^{\lambda'\lambda}\delta_{\vec{k}'\vec{k}}\, ,\\
\label{aadagger}
           \lbrack \hat{a}_{\vec{k}'}^{(\lambda')},
\hat{a}_{\vec{k}}^{(\lambda)}\rbrack &=& \lbrack \hat{a}_{\vec{k}'}^{(\lambda')\dagger},
\hat{a}_{\vec{k}}^{(\lambda)\dagger}\rbrack = 0\, .
\label{aa}
\end{eqnarray}
\end{mathletters}
Hence, $\hat{a}_{\vec{k}}^{(\lambda)\dagger}$ creates and
$\hat{a}_{\vec{k}}^{(\lambda)}$ annihilates a graviton with wave vector
$\vec{k}$ and polarization $\lambda$. More precisely, they operate into the 
vacuum $\vert 0\rangle$ such that
\begin{mathletters}
\label{operations}
\begin{eqnarray}
\hat{a}_{\vec{k}}^{(\lambda)\dagger}\vert 0\rangle &=& \sqrt{16\pi
G\hbar}\vert 1\rangle\, ,\\
\label{creation}
            \hat{a}_{\vec{k}}^{(\lambda)}\vert 0\rangle &=& 0\, ,
\label{annihilation}
\end{eqnarray}
\end{mathletters}
where $\vert 1\rangle$ is a one-graviton state. If we put everything here
derived in together, we find that the operator $\hat{H}_{int}$ of Eq.~(\ref{Hinter}) takes
the form:
\begin{equation}
\hat{H}_{int} = {1\over{2\mu}}\sum_{\vec{k}}\sum_{\lambda=1}^2
(2L^3\omega)^{-1/2}\epsilon^{(\lambda)ab}\cos({1\over
2}\vec{k}\cdot\vec{r})(e^{-i\omega t}\hat{a}^{(\lambda)}_{\vec{k}} +
e^{i\omega t}\hat{a}_{\vec{k}}^{(\lambda)\dagger})\hat{p}_a\hat{p}_b\, .
\label{Hint}
\end{equation}

        Now, the Golden Rule~\cite{Messiah} implies that if in a transition  from the state $\vert
i\rangle$ to the state $\vert f\rangle$ a graviton with energy $\hbar\omega$,
wave vector $\vec{k}$ and polarization $\lambda$ is emitted, the corresponding
transition rate is
\begin{equation}
\R_{i\rightarrow f} = {{2\pi}\over\hbar} \vert\langle f\vert
{1\over{2\mu}}(2L^3\omega)^{-1/2}\epsilon^{(\lambda)ab}
\cos({1\over 2}\vec{k}\cdot\vec{r})\hat{a}^{(\lambda)\dagger}_{\vec{k}}
\hat{p}_a\hat{p}_b\vert i\rangle\vert^2\rho(E_f)\, ,
\label{transitrate2}
\end{equation}
where
\begin{equation}
\rho(E_f) := {{\omega^2 L^3}\over{8\pi^3\hbar}}\,d\Omega\
\label{density}
\end{equation}
is the density of states close to the final state, and $d\Omega$ is the solid
angle into which the graviton emerges. If the state $\vert i\rangle$ represents
a zero-graviton state and the state $\vert f\rangle$ a one-graviton state, Eq.(A20)
implies
\begin{equation}
\R_{i\rightarrow f} = {{G\omega}\over{2\pi\hbar\mu^2}} \vert\langle f\vert\cos({1\over
2}\vec{k}\cdot\vec{r})\epsilon^{(\lambda)ab}\hat{p}_a\hat{p}_b\vert
i\rangle\vert^2\,d\Omega\
\label{transitrate3}
\end{equation}
or, in SI units:
\begin{equation}
\R_{i\rightarrow f} = {{G\omega}\over{2\pi c^5\hbar\mu^2}} \vert\langle f\vert\cos({1\over
2}\vec{k}\cdot\vec{r})\epsilon^{(\lambda)ab}\hat{p}_a\hat{p}_b\vert
i\rangle\vert^2\,d\Omega\, ,
\label{transitrate4}
\end{equation}
which is the same as in Eq.~(\ref{transitrate'}).

\section{Integration Over All Directions of the Gravitational Radiation}
\label{app:B}

In this Appendix we shall calculate the integrated transition rate by integrating over all the possible diresctions into which the graviton may emerge. To evaluate the integral, we assume first that all the gravitons are the so called $h^{12}$-gravitons propagating to $z$-direction. Then the only non-zero components of the polarization tensor $\epsilon ^{ab}$ are
\begin{equation}
\label{poltens}
\epsilon ^{12} = \epsilon ^{21} = 1\, .
\end{equation}

Now, the idea is here that the components of the polarization tensor change when the direction of propagation of the emitting graviton changes. When the gravitons no more propagate to $z$-direction, we introduce a new Cartesian coordinate system $K'$ which is rotated such that the gravitons always propagate,  in frame $K'$, along $z$-axis, and the original coordinate system $K$ is kept fixed.  The relationship between the orthonormal  basis vectors $\hat e _a$ and $\hat e' _a$ of the frames $K$ and $K'$ is:
\begin{mathletters}
\begin{eqnarray}
\label{basisvec}
\hat e'_1 &=& \sin \phi ~\hat e_1 + \cos \phi ~\hat e_2\, ,\\
\hat e'_2 &=& - \cos \theta \cos \phi ~\hat e_1 +\cos \theta \sin \phi ~\hat e_2 + \sin \theta ~\hat e_3\, ,\\
\hat e'_3 &=& \sin \theta \cos \phi ~\hat e_1 + \sin \theta \sin \phi ~\hat e_2 + \cos \theta ~\hat e_3\, .
\end{eqnarray}
\end{mathletters}
where $\theta$ and $\phi$ are spherical angles. These relations imply that the polarization tensor corresponding to  $h^{12}$-gravitons propagating to arbitrary direction is, in frame $K$, represented by a matrix:
\begin{equation}
\epsilon =
\left(\begin{array}{ccc}
 -\cos \theta \sin 2\phi & \cos \theta (2\sin ^2 \phi -1)  & \sin \theta \sin \phi   \\
\cos \theta (2\sin ^2 \phi -1)& \cos \theta \sin 2\phi & \sin \theta \cos \phi \\
\sin \theta \sin \phi & \sin \theta \cos \phi & 0
\label{polmatrix}
\end{array}\right)
\end{equation}
To begin with we perform the summation in Eq.~(\ref{transitrate}). Because the momentum operators commute with each other, we get:
\begin{eqnarray}
\int\vert \langle f \vert \epsilon^{ab}\hat p_a \hat p_b \vert i \rangle \vert ^2 d\Omega &=& 
\int\vert
\epsilon^{11} \langle f \vert \hat p_1 ^2 \vert i \rangle +
\epsilon^{22} \langle f \vert \hat p_2 ^2 \vert i \rangle +
\epsilon^{33} \langle f \vert \hat p_3 ^2 \vert i \rangle +
2\epsilon^{12} \langle f \vert \hat p_1 \hat p_2 \vert i \rangle \nonumber \\
& &+~2\epsilon^{13} \langle f \vert \hat p_1 \hat p_3 \vert i \rangle +
2\epsilon^{23} \langle f \vert \hat p_2 \hat p_3 \vert i \rangle
\vert ^2 d\Omega
\label{aftersum}
\end{eqnarray}
After squaring and integrating the expression (\ref{aftersum}) over all directions, we have:
\begin{eqnarray}
\int\vert \langle f \vert \epsilon^{ab}\hat p_a \hat p_b \vert i \rangle \vert ^2 d\Omega &=& 
\frac{2\pi}{3}\Bigl( \vert \langle f \vert \hat p_1 ^2 \vert i \rangle \vert ^2+ \vert \langle f \vert \hat p_2 ^2 \vert i \rangle \vert ^2 +4\vert \langle f \vert \hat p_1 \hat p_2 \vert i \rangle \vert ^2+8\vert \langle f \vert \hat p_1 \hat p_3 \vert i \rangle \vert ^2 \nonumber \\
& & +~8\vert \langle f \vert \hat p_2 \hat p_3 \vert i \rangle \vert ^2 - \langle f \vert \hat p_2 ^2 \vert i \rangle^\ast\langle f \vert \hat p_1 ^2 \vert i \rangle - \langle f \vert \hat p_1 ^2 \vert i \rangle^\ast\langle f \vert \hat p_2 ^2 \vert i \rangle\nonumber \\
& & +~8 \langle f \vert \hat p_1 \hat p_3 \vert i \rangle^\ast\langle f \vert \hat p_2 \hat p_3 \vert i \rangle + 8 \langle f \vert \hat p_2 \hat p_3 \vert i \rangle^\ast\langle f \vert \hat p_1 \hat p_3 \vert i \rangle \Bigr ) \, .
\label{afterint}
\end{eqnarray}
In Appendix \ref{app:D} we shall calculate the transition amplitudes above in all details. If we now use the results~(\ref{otheramps}), we get for the integrated transition rate a simple and nice expression, by taking into account that gravitons have two independent polarization states:
\begin{equation}
\label{transitratesimp}
R_{i\rightarrow  f} = \frac{16 \omega}{3 M^2}\vert \langle f\vert \hat p_1 ^2 \vert i \rangle \vert ^2 \, ,
\end{equation}
in the units where $G=c=\hbar=1$.

\section{G1 Selection Rules}
\label{app:C}

In this Appendix we shall consider, in the lowest order approximation, the selection rules for the spontaneous emissions and absorptions of gravitons by the microscopic black hole pair. These selection rules we shall call G1 selection rules, in analogy to the E1 selection rules in atomic physics. The most convienient way of deriving these rules is to use the so called Wigner--Eckart theorem~\cite{Messiah}.  That theorem concerns the irreducible tensor operators in the spherical basis. The Cartesian components $\hat p_{1,~2,~3}$ of the momentum operator can be written in terms of the standard components $\hat p_{-1,~+1,~0} $ of the irreducible momentum operator $\hat{\vec p}$ of rank 1 as:
\begin{mathletters}
\begin{eqnarray}
\hat p_1 &=& -\frac{1}{\sqrt{2}}\left(\hat p_{+1}-\hat p_{-1}\right) \, ,  \\
\hat p_2 &=& \frac{i}{\sqrt{2}}\left(\hat p_{+1}+\hat p_{-1}\right) \, ,  \\
\hat p_3 &=& \hat p_0 \, .
\label{irreduop}
\end{eqnarray}
\end{mathletters}
To find the G1 selection rules, we must find all  the allowed changes in the angular momentum of the microscopic black hole pair between the initial and final states. To know these changes one has to consider the transition amplitude given in Eq.~(\ref{transitratesimp}).  The transition amplitude can be written in terms of the standard components as:
\begin{equation}
\langle f \vert  \hat p_1^2 \vert i \rangle = \langle f \vert \frac{1}{2}\left(  
\hat p_{+1}^2-\hat p_{+1} \hat p_{-1} - \hat p_{-1} \hat p_{+1} + \hat p_{-1}^2 \right) \vert i \rangle \, ,
\label{irreduamp1}
\end{equation}
where all the products between the standard components can be written, by using the definition of the tensor product, as:
\begin{eqnarray}
\langle f \vert  \hat p_1^2 \vert i \rangle &=&  \frac{1}{2}\Bigl[  \langle f \vert \sum_{J, M} (1,1,1,1\vert J,M)[{\hat {\vec p}} {\hat {\vec p}}]_{JM}  \vert i \rangle - \langle f \vert \sum_{J, M} (1,1,1,\!-\!1\vert J,M)[{\hat {\vec p}} {\hat {\vec p}}]_{JM}  \vert i \rangle- \nonumber \\
& &~\langle f \vert \sum_{J, M} (1,\!-\!1,1,1\vert J,M)[{\hat {\vec p}} {\hat {\vec p}}]_{JM}  \vert i \rangle + \langle f \vert \sum_{J, M} (1,\!-\!1,1,\!-\!1\vert J,M)[{\hat {\vec p}} {\hat {\vec p}}]_{JM}  \vert i \rangle   \Bigr] \, ,
\label{irreduamp2}
\end{eqnarray} 
where $[{\hat {\vec p}} {\hat {\vec p}}]_{JM}$ denotes the $JM$-component of the irreducible tensor of rank 2, $J$ corresponds to the eigenvalue of the total angular momentum operator $\hat J$, and $M$ is the corresponding eigenvalue of the projection operator $\hat J_z$. After taking the sum over  $J$ and $M$ Eq.~(\ref{irreduamp2}) reduces to 
\begin{equation}
\langle f \vert  \hat p_1^2 \vert i \rangle = \frac{1}{2}\langle f \vert [{\hat {\vec p}} {\hat {\vec p}}]_{22}  \vert i \rangle -
\frac{1}{\sqrt{3}}\langle f \vert [{\hat {\vec p}} {\hat {\vec p}}]_{00}  \vert i \rangle - \frac{6}{\sqrt{5}}\langle f \vert [{\hat {\vec p}} {\hat {\vec p}}]_{20}  \vert i \rangle + \frac{1}{2}\langle f \vert [{\hat {\vec p}} {\hat {\vec p}}]_{2~-\!2}  \vert i \rangle \, ,
\label{irreduamp3}
\end{equation}
which, by Wigner-Eckart theorem, is equal to:
\begin{eqnarray}
\langle f \vert  \hat p_1^2 \vert i \rangle &=& (-1)^{l_f-m_f}\Biggl[ 
\frac{1}{2}
\left(\begin{array}{ccc}
l_f&2&l_i \\
-m_f&2&m_i
\end{array}\right)
 \left( n_f~l_f \vert\vert [{\hat {\vec p}} {\hat {\vec p}}]_{22}  \vert\vert  n_i~l_i \right)  -\nonumber \\
& &~\frac{1}{\sqrt{3}}
\left(\begin{array}{ccc}
l_f&0&l_i \\
-m_f&0&m_i
\end{array}\right)
 \left( n_f~l_f \vert\vert [{\hat {\vec p}} {\hat {\vec p}}]_{00}  \vert\vert  n_i~l_i \right)-  \nonumber \\
& &~ \frac{6}{\sqrt{5}}
\left(\begin{array}{ccc}
l_f&2&l_i \\
-m_f&0&m_i
\end{array}\right)
 \left( n_f~l_f \vert\vert [{\hat {\vec p}} {\hat {\vec p}}]_{20}  \vert\vert  n_i~l_i \right) +  \nonumber \\
& &~ \frac{1}{2}
\left(\begin{array}{ccc}
l_f&2&l_i \\
-m_f&-2&m_i
\end{array}\right)
\left( n_f~l_f \vert\vert [{\hat {\vec p}} {\hat {\vec p}}]_{2~-\!2} \vert\vert  n_i~l_i \right) \Biggr] \, ,
\label{WE}
\end{eqnarray}
where $m_i$ and $m_f$ are the eigenvalues of the $z$-component of the angular momentum operator at the final and at the initial states of the microscopic black hole pair, $\left(\begin{array}{ccc}l_f&L&l_i \\ m_f&M&m_i \end{array} \right)$ is a 3j-symbol  and $(f\vert\vert \hat T_L \vert\vert i)$ is the reduced matrix element of the irreducible tensor operator of rank $L$. 

It is well known that the 3j-symbol vanishes unless the following two conditions hold for the eigenvalues of the angular momenta:
\begin{mathletters}
\begin{eqnarray}
\vert l_f - L \vert \leq l_i \leq&l_f&+L \, \\ 
m_i-m_f+M&=&0 \, .
\label{conds}
\end{eqnarray}
\end{mathletters}
These conditions, together with Eq.~(\ref{WE}), imply that the allowed transitions are the ones where $\Delta l = 0,~\pm 1,~\pm 2$ and $\Delta m_l= 0,~\pm 2$. However, there is a further constraint which comes from the parity conservation: Since the position operator $\hat x^a$ is odd in reflections for every $a=1,2,3$, $\hat p^a$ is also odd for all $a$. Therefore $\hat p^a\hat p^b$ must behave as an even linear operator in reflections, and the states of the microscopic black hole pair must not change parity in G1 transitions. Since the parity of any state is given by $(-1)^l$, the value of $l$ cannot be $\pm 1$. As the final G1 selection rules for the microscopic black hole pairs  we find that the only allowed transitions in the lowest order approximation are the ones where
\begin{mathletters}
\label{G1}
\begin{eqnarray}
\Delta l = 0,~\pm 2 \, , \\
\Delta m_l = 0,~\pm 2 \, .
\end{eqnarray}
\end{mathletters}

\section{Transition Rates}
\label{app:D}

In this Appendix we shall derive explicit expressions for the transition rates in G1 transitions in the microscopic black hole pairs. In addition, we shall estimate the transition rates numerically in some well-chosen cases. Furthermore, we shall estimate the life time of the initial quantum state of the black hole pair.

In a position representation the initial and the final quantum states, $\vert i \rangle$ and $\vert f \rangle$, respectively, are represented by the "hydrogenic wave functions", where the factor $q^2/4\pi\epsilon_0$ is replaced by $GM^2$, where $G$ is the gravitational constant and $M$ is the mass of each hole~\cite{Arfken}:
\begin{eqnarray}
\Psi _{nlm_l}(r,\theta,\phi )&=& \left[
\left(\frac{GM^3}{n\hbar^2}\right)^3\frac{(n-l-1)!}{2n(n+l)!}
\right]^{1/2}
\exp\left(\frac{-GM^3}{2n\hbar^2}r\right)
\left(\frac{GM^3}{n\hbar^2}r\right)^l \times \nonumber \\
& &~L_{n-l-1}^{2l+1}\left(\frac{GM^3}{n\hbar^2}r\right)
Y_{lm_l}\left(\theta ,\phi\right)\, ,
\label{wavefunc}
\end{eqnarray}
where
\begin{equation}
L_k^s (x) := \sum_{m=0}^s(-1)^m\frac{(s+k)!}{(s-m)!(k+m)!m!}x^m
\label{laguerre}
\end{equation}
is the associated Laguerre polynomial, and 
\begin{equation}
Y_{lm_l}(\theta ,\phi ):= (-1)^{m_l}\left[ \frac{(2l+1)(l-m_l)!}{4\pi(l+m_l)!} \right]^{1/2}P_l^{m_l}(\cos\theta ) \exp (im_l\phi )
\label{ylm}
\end{equation}
is the spherical harmonic function. $P_l^{m_l}(x)$ is the associated Legendre polynomial. When calculating an expression for the transition rate, we will mostly be interested in the transitions where the initial and the final states are in their maximal projection state where $m_ {l_i} =l_i$ and $m_{l_f}=l_f$. In that case we may use the result:
\begin{equation}
P_l^l(\cos\theta ) = (2l-1)!!\sin^l\theta\, ,
\label{Plm}
\end{equation}
where the double factorial is defined as $n!! := n(n-2)(n-4)\cdot \cdots$ and $0!!=1$.

In Appendix~\ref{app:C} we derived the G1 selection rules~(\ref{G1}).  Because of these rules we must consider four different cases: We first calculate the three transition rates $R_{i\rightarrow f}$ where either $\Delta l= 0$ or $\Delta l=\pm 2$ such that the initial and final states are maximal projection states, and then we consider the case where $\Delta l = 0$ and $\Delta m_l = -2$. 

\subsection{$\Delta l =-2$}

We first calculate the transition rate in a case where $\Delta l = -2 = \Delta m_l$. In other words we take the initial and final states to be $\vert i \rangle = \vert n,l,l\rangle$ and $\vert f \rangle = \vert n-k, l-2, l-2\rangle$ $(k=1,2,3,\dots)$ such that $l > 2$ and $n \gg 1$.  We can write these states in the position representation using Eq.~(\ref{wavefunc}). To evaluate the integrals in Eq.~(\ref{afterint}), we must express the momentum operators $\hat p_a$ in the spherical coordinates:
\begin{mathletters}
\label{momenta}
\begin{eqnarray}
\hat p_1 &=& -i \hbar \left( \sin\theta\cos\phi\frac{\partial }{\partial r}+ \frac{\cos\theta\cos\phi}{r}\frac{\partial}{\partial\theta}- 
\frac{\sin\phi}{r\sin\theta}\frac{\partial}{\partial\phi}\right) \, ,\\
\hat p_2 &=& -i \hbar \left( \sin\theta\sin\phi\frac{\partial }{\partial r}+ \frac{\cos\theta\sin\phi}{r}\frac{\partial}{\partial\theta}- 
\frac{\cos\phi}{r\sin\theta}\frac{\partial}{\partial\phi}\right) \, ,\\
\hat p_3 &=& -i \hbar \left( \cos \theta \frac{\partial}{\partial r}-\frac{\sin\theta}{r}\frac{\partial}{\partial \theta}\right) \, .
\end{eqnarray}
\end{mathletters}
After employing the expressions for the wave functions and for the momentum operators one gets a rather messy expression for the first transition amplitude in Eq.~(\ref{afterint}), which we still have to integrate (in the units where $c=G=\hbar=M=1$):
\begin{eqnarray} 
\label{transit}
\langle f\vert \hat p_1^2 \vert i\rangle &=& \langle \hat p_1^\dagger f \vert \hat p_1 i\rangle  
				=  \langle \hat p_1 f \vert \hat p_1 i\rangle  \nonumber \\
&=& \frac{(2l-1)!!(2l-5)!!}{8\pi n^{l+2}(n-k)^l}\left[\frac{(2l+1)(2l-3)(n-l-1)!(n-k-l+1)!}{(n+l)!(2l)!(n-k+l-2)!(2l-4)!} \right]^{1/2} \times \nonumber \\
& &~\sum_{m=0}^{n-l-1}a_m\sum_{m'=0}^{n-k-l+1}b_{m'}\int_0^\infty dr r^2 \int_0^\pi d\theta \sin\theta \int_0^{2\pi} d\phi \exp (i2\phi) \times  \nonumber \\
& &~\Bigl[ \sin^{2l}\theta \cos^2\phi R'_i(r) R'_f(r)+l\sin^{2l-2}\theta\cos^2\theta\cos^2\phi\frac{1}{r} R_i(r) R'_f(r)- \nonumber \\
& &~\frac{il}{2r}\sin^{2l-2}\theta\sin 2\phi  R_i(r) R'_f(r) +  \frac{l-2}{r}\sin^{2l-4}\theta\cos^2\theta\cos^2\phi R'_i(r) R_f(r)+ \nonumber \\
& &~\frac{l(l-2)}{r^2}\sin^{2l-4}\theta\cos^4\theta\cos^2\phi R_i(r) R_f(r)-\frac{i(l-2)}{2r}\sin^{2l-2}\theta\sin 2\phi  R'_i(r) R_f(r)+\nonumber \\
& &~\frac{l(l-2)}{r^2}\sin^{2l-4}\theta\sin^2\phi R_i(r) R_f(r) \Bigr] \, ,
\end{eqnarray}
where we have denoted:
\begin{mathletters}
\begin{eqnarray} 
\label{defs}
a_m&:=& (-1)^m\frac{(n+l)!}{(n-l-m-1)!(2l+m+1)!m!n^m} \, ,\\
b_{m'}&:=& (-1)^{m'}\frac{(n-k+l-2)!}{(n-k-l-m'+1)!(2l+m'-3)!m'!(n-k)^{m'}} \, ,\\
R_i(r)&:=& r^{l+m}\exp \left(-\frac{r}{2n}\right) \, ,\\
R_f(r)&:=& r^{l+m'-2}\exp \left(-\frac{r}{2(n-k)}\right) \, ,
\end{eqnarray}
\end{mathletters}
and the prime $'$ denotes the partial derivative with respect to $r$.
The $\phi$-part of the integral is very easy to perform, and the radial part gives four different integrals which can be integrated separately by using the well-known result: 
\begin{equation}
\int_0^\infty dx x^n \exp (-ax) = \frac{n!}{a^{n+1}} \, .
\label{radial}
\end{equation}
Moreover, it is straightforward to show that the $\theta$-integrals in Eq.~(\ref{transit}) can be reduced to the following integral:
\begin{equation} 
\label{theta}
\int_0^\pi d\theta \sin^{2l+1}\theta = \frac{2^{2l+1}l!^2}{(2l+1)!} \, .
\end{equation}
When substituting all these integrals into Eq.~(\ref{transit}), we get, for the transitions where $\Delta l=-2$ such that the system always remains in its maximal projection state:
\begin{eqnarray} 
\label{transit1}
\langle f\vert \hat p_1^2 \vert i\rangle &=& \frac{1}{8 n^{l+2}(n-k)^l} \left[\frac{2l(2l-2)(n+l)!(n-k+l-2)!(n-l-1)!(n-k-l+1)!}{(2l+1)(2l-1)} \right]^{1/2} \times \nonumber \\
& &~\sum_{m=0}^{n-l-1}\frac{(-1)^m }{(n-l-m-1)!(2l+m+1)!m!n^m}  \times  \nonumber \\ 
& &~\sum_{m'=0}^{n-k-l+1}\frac{(-1)^{m'}}{(n-k-l-m'+1)!(2l+m'-3)!m'!(n-k)^{m'}} \times \nonumber \\
& &~\left[ \frac{2n(n-k)}{2n-k}^{2l+m+m'}\right]
\Biggl\lbrace (2l+m+m'-2)!(2l+m+2)m'\frac{2n-k}{2n(n-k)} - \nonumber \\
& &~(2l+m+m'-1)! \left( \frac{2l+m'-4}{2n} + \frac{2l+m-1}{2(n-k)}\right)+\frac{(2l+m+m')!}{2(2n-k)}\Biggl\rbrace\, .
\end{eqnarray}
In a very similar manner one can show that the rest of the transition amplitudes in Eq.~(\ref{afterint}) in this particular case, are given by:
\begin{mathletters}
\label{otheramps}
\begin{eqnarray} 
\langle f\vert \hat p_2^2 \vert i\rangle &=&- \langle f\vert \hat p_1^2 \vert i\rangle \, \\
\langle f\vert \hat p_j \hat p_3 \vert i\rangle &=&0  ~ ~\forall j=1, 2, 3 \, , \\
\langle f\vert \hat p_1 \hat p_2 \vert i\rangle &=& i\langle f\vert \hat p_1^2 \vert i\rangle \, . 
\end{eqnarray}
\end{mathletters}
The only difference between the rest of the amplitudes in Eq.~(\ref{afterint}) and the first amplitude~(\ref{transit1}) comes from the $\phi$-integration. 

The angular frequency $\omega$ depends on the quantum number $n$ and $\Delta n =-k$ as follows:
\begin{equation}
\omega = -\frac{1}{2}\left( \frac{1}{n^2}-\frac{1}{(n-k)^2}\right)\, .
\label{w}
\end{equation}
As a final result, when the $n$-dependence of the angular frequency is taken into account, the transition rate for the microscopic black hole pairs becomes, in the case where $\Delta l =-2 =\Delta m_l$:
\begin{eqnarray} 
\label{Rif1}
R_{i\rightarrow f} &=&-  \frac{1}{24}\left(\frac{1}{n^2}-\frac{1}{(n-k)^2}\right) \frac{2l(2l-2)(n+l)!(n-k+l-2)!(n-l-1)!(n-k-l+1)!}{(2l+1)(2l-1) n^{2l+4}(n-k)^{2l}} \times \nonumber \\
& &~\Biggl\lbrace\sum_{m=0}^{n-l-1}\frac{(-1)^m}{(n-l-m-1)!(2l+m+1)!m!n^m}  \times  \nonumber \\ 
& &~\sum_{m'=0}^{n-k-l+1}\frac{(-1)^{m'}}{(n-k-l-m'+1)!(2l+m'-3)!m'!(n-k)^{m'}} \times \nonumber \\
& &~\left[ \frac{2n(n-k)}{2n-k}^{2l+m+m'}\right]
\Biggl[ (2l+m+m'-2)!(2l+m+2)m'\frac{2n-k}{2n(n-k)} - \nonumber \\
& &~(2l+m+m'-1)! \left( \frac{2l+m'-4}{2n} + \frac{2l+m-1}{2(n-k)}\right)+\frac{(2l+m+m')!}{2(2n-k)}\Biggl]\Biggl\rbrace^2\, .
\end{eqnarray}
If one prefers SI-units to the natural units, then the above transition rate should be multiplied by the factor $\frac{G^7M^{15}}{c^5\hbar^8}$. 

As an example, let us now consider the binary black hole system consisting of two microscopic black holes  with equal masses $M=M_{Pl}\approx 22 \mu$g. In that case the transition rate between two neighbouring states from the state $\vert i\rangle = \vert 100, 50, 50\rangle$ to the state $\vert f\rangle = \vert 99, 48, 48\rangle$ is approximately:
\begin{equation} 
\label{numRif}
R_{i\rightarrow f} \approx 2.25\times10^{27}~\frac{1}{\mathrm{s}} 
\end{equation}
and the corresponding life time of the initial state in this transition is 
\begin{equation} 
\label{numtau}
\tau = \frac{1}{R_{i\rightarrow f}}\approx 4.46\times10^{-28}~\mathrm{s}\, . 
\end{equation}

When $\Delta m_l \neq -2$, we may calculate transition rates from Eq.~(\ref{WE}), since the reduced matrix elements do not depend on the value of the $z$-component of the angular momentum operator. If $\Delta m_l=+2$ one can easily show that the transition rate is given by 
\begin{equation}
\tilde R_{i\rightarrow f} = \left(\begin{array}{ccc}
l_f&2&l_i \\
-\tilde m_f&2&\tilde m_i
\end{array}\right)^2
\left(\begin{array}{ccc}
l_f&2&l_i \\
-m_f&-2&m_i
\end{array}\right)^{-2}R_{i\rightarrow f}\, ,
\label{ml+2}
\end{equation}
where $R_{i\rightarrow f}$ is the transition rate related to the case $\Delta m_l=-2$ and the tilde~$\tilde{}$ denotes the case $\Delta m_l=+2$. 

In the same way one can show that the case  $\Delta m_l =0$ is related to the case  $\Delta m_l =-2$ such that:
\begin{equation}
\tilde R_{i\rightarrow f} = \left(\begin{array}{ccc}
l_f&2&l_i \\
-\tilde m_f&0&\tilde m_i
\end{array}\right)^2
\left(\begin{array}{ccc}
l_f&2&l_i \\
-m_f&-2&m_i
\end{array}\right)^{-2}R_{i\rightarrow f}\, ,
\label{ml+2}
\end{equation}
where $R_{i\rightarrow f}$ is the transition rate related to the case $\Delta m_l=-2$ and now the tilde~ $\tilde{}$ denotes the case $\Delta m_l=0$. 

\subsection{$\Delta l = 0$}

Let us next consider the case where $\Delta l=0$ such that the system always stays in its maximal projection state. In other words the initial state is $\vert i \rangle = \vert n,l,l\rangle$ and the final state is $\vert f \rangle = \vert n-k, l, l\rangle$. The position representation of these states can be read from Eq.~(\ref{wavefunc}). In fact, the whole procedure to derive an expression for the transition rate in this case is analogous to the previous case where $\Delta l = -2$. Therefore we just give the final result:
\begin{eqnarray} 
\label{Rif2}
R_{i\rightarrow f} &=&-  \frac{1}{3}\left(\frac{1}{n^2}-\frac{1}{(n-k)^2}\right) \left(\frac{l+1}{2l+3}\right)^2\frac{(n+l)!(n-k+l)!(n-l-1)!(n-k-l-1)!}{n^{2l+4}(n-k)^{2l+4}} \times \nonumber \\
& &~\Biggl\lbrace\sum_{m=0}^{n-l-1}\frac{ (-1)^m }{(n-l-m-1)!(2l+m+1)!m!n^m}  \times  \nonumber \\ 
& &~\sum_{m'=0}^{n-k-l-1}\frac{(-1)^{m'} }{(n-k-l-m'-1)!(2l+m'+1)!m'!(n-k)^{m'}} \times \nonumber \\
& &~\left[ \frac{2n(n-k)}{2n-k}^{2l+m+m'+2}\right]
\Biggl[ (2l+m+m')! \Biggl( (l(m+m'+l) +mm')- \nonumber \\
& &~ \frac{6l^3-4l^2+3l +(m+m')(2l^2-4l)}{2l+2}\Biggr)  \frac{2n-k}{2n(n-k)} + \nonumber \\
& &~(2l+m+m'+1)! \left( \frac{l-m'+1}{2n} - \frac{l+m-1}{2(n-k)}\right)+\frac{(2l+m+m'+2)!(n-k)}{2n-k}\Biggl]\Biggl\rbrace^2\, .
\end{eqnarray}
For instance, if the initial and the final states are fixed such that  the initial state is $\vert i \rangle = \vert 100,50,50\rangle$ and the final state is $\vert f \rangle = \vert 99,50,50\rangle$, then the transition rate between these states is
\begin{equation} 
\label{numRif2}
R_{i\rightarrow f} \approx 1.44\times10^{28}~\frac{1}{\mathrm{s}} 
\end{equation}
and the corresponding life time of the initial state in this transition is 
\begin{equation} 
\label{numtau}
\tau \approx 6.94\times10^{-29}~\mathrm{s}\, . 
\end{equation}

When $\Delta m_l \neq 0$, we cannot use Eq.~(\ref{WE}) to derive expressions for the transition rates; instead we have to perform another lengthy calculation. The difference between these cases comes from the definition of the associated Legendre polynomial, since we cannot use Eq.~(\ref{Plm}), which  holds only for the maximal projection states, in the cases where  $\Delta m_l \neq 0$.  Because our system is not necessarily in a maximal projection state we have to use the general definition for the associated Legendre polynomials:
\begin{equation}
P_l^{m_l} (x) := \frac{(1-x^2)^{m_l/2}}{2^ll!}\sum_{i=0}^l\frac{(-1)^{l-i}l!}{i!(l-i)!}\prod_{j=0}^{l+m_l-1}\vert 2i-j \vert 
x^{2i-l-m_l} \, .
\label{genPlm}
\end{equation}
As this definition comes to effect in Eq.~(\ref{wavefunc}), the transition rate still is given by~(\ref{transitratesimp}). One can show, after performing some integrals, that when $\Delta l = 0$ and $\Delta m_l =-2$, the transition rates of our system obey, for large $n$,  the following expression:
\begin{eqnarray}
\label{Rif3}
R_{i\rightarrow f} &=&-  \frac{1}{24}\left(\frac{1}{n^2}-\frac{1}{(n-k)^2}\right) \frac{(2l+1)^2 2^{2m_l}m_l!^2 (l-m_l)!(l-m_l+2)!(n+l)!}{n^{2l+4}(n-k)^{2l+4}(l+m_l)!(l+m_l-2)!}  \times \nonumber \\ 
& &~(n-k+l)!(n-l-1)!(n-k-l-1)! \Biggl\lbrace\sum_{m=0}^{n-l-1}  \sum_{m'=0}^{n-k-l-1} 
\sum_{s=0}^l  \sum_{s'=0}^l  a_m b_{m'}  \times \nonumber \\
& &~\Biggl[ I_1K_{ss'}T_{ss'}^1+I_3\Biggl( -K_{ss'}T_{ss'}^2+K_{ss'}T_{ss'}^3\Biggr) + 
I_2\Biggl(- K_{ss'}T_{ss'}^4+K_{ss'}T_{ss'}^5\Biggr) +\nonumber \\ 
& &~ I_4\Biggl( K_{ss'}T_{ss'}^6-K_{ss'}T_{ss'}^7 +K_{ss'}T_{ss'}^8 \Biggr) \Biggr] \Biggl\rbrace^2\, ,
\end{eqnarray} 
where we have denoted
\begin{mathletters}
\begin{eqnarray} 
\label{defs}
a_m&:=& \frac{(-1)^m}{(n-l-m-1)!(2l+m+1)!m!n^m} \, ,\\
b_{m'}&:=& \frac{(-1)^{m'}}{(n-k-l-m'-1)!(2l+m'+1)!m'!(n-k)^{m'}} \, ,\\
I_1&:=&\int_0^\infty dr r^2 R'_i(r)R'_f(r) \, ,\\
I_2&:=&\int_0^\infty dr r R'_i(r)R_f(r) \, ,\\
I_3&:=&\int_0^\infty dr r R_i(r)R'_f(r) \, ,\\
I_4&:=&\int_0^\infty dr  R_i(r)R_f(r) \, ,\\
R_i(r)&:=& r^{l+m}\exp \left(-\frac{r}{2n}\right) \, ,\\
R_f(r)&:=& r^{l+m'}\exp \left(-\frac{r}{2(n-k)}\right) \, , \\
 K_{ss'}&:=& \frac{(-1)^{2l-s-s'}}{2^{2l}s!(l-s)!s'!(l-s')!} \prod_{i=0}^{l+m_l-1}\vert 2s-i \vert \prod_{j=0}^{l+m_l-3}\vert 2s'-j \vert \, ,\\
T_{ss'}^1&:=& \frac{2^{s+s'-l}(s+s'-l)!}{(2s+2s'-2l+3)(2s+2s'-2l+1)!} \, ,\\
T_{ss'}^2&:=& \frac{2^{s+s'-l}(s+s'-l)!}{(2s+2s'-2l+1)!}\left( \frac{1}{2(2s+2s'-2l+3)} +2s'-m_l-l \right) \, ,\\
T_{ss'}^3&:=& \frac{m_l(2m_l)!}{2^{s+s'-l+2m_l}(s+s'-l-1)!(2m_l-1)m_l!^2} \, ,\\
T_{ss'}^4&:=& \frac{2^{s+s'-l}(s+s'-l)!}{2m_l(2m_l-2)(2s+2s'-2l)!}\left( \frac{2-m_l}{2m_l-4} +2s'-m_l-l+2 \right) \, ,\\
T_{ss'}^5&:=& \frac{(2-m_l)(2m_l)!2^{s+s'-l-2m_l}(s+s'-l)!}{m_l!^2(2s+2s'-2l+1)!} \, ,\\
T_{ss'}^6&:=& \frac{2^{s+s'-l}(s+s'-l)!}{2(2s+2s'-2l)!} \Biggl( \frac{m_l-2}{(2s+2s'-2l+1)(2m_l-2)} + \nonumber \\ 
& ~&~\frac{2s'-l-m_l+2}{(2s+2s'-2l+3)(2s+2s'-2l+1)} - \frac{(m_l-2)(2s-l-m_l)}{m_l(2m_l-2)} + \nonumber \\ & ~&~\frac{(2s-l-m_l)(2s'-l-m_l+2)}{m_l(2s+2s'-2l+1)}  \Biggr) \, ,\\
T_{ss'}^7&:=& \frac{(2m_l)!2^{s+s'-l-2m_l}(s+s'-l)!}{m_l!^2} \Biggl(
 \frac{m_l(2-m_l)}{(2s+2s'-2l)!(2m_l-5)(2m_l-3)(2m_l-1)} + \nonumber \\ 
& ~&~\frac{(2-m_l)(2s-l-m_l)}{(s+s'-l+)!^22^{2s+2s'-2l}} + \frac{m_l(2s'-l-m_l+2)}{(2m_l-3)(2m_l-1)(2s+2s'-2l)!} + \nonumber \\ & ~&~\frac{m_l(2-m_l)}{(2m_l-1)2^{2s+2s'-2l}(s+s'-l)!^2(2s+2s'-2l+2)}  \Biggr) \, ,\\
T_{ss'}^8&:=& \frac{(2-m_l)2^{s+s'-l}(s+s'-l)!}{2(2s+2s'-2l+1)!} \, .
\end{eqnarray}
\end{mathletters}
When $\Delta m_l = +2$, we may use Eq.~(\ref{WE}) to give an expression for the transition rate
\begin{equation}
\tilde R_{i\rightarrow f} = \left(\begin{array}{ccc}
l_f&2&l_i \\
-\tilde m_f&2&\tilde m_i
\end{array}\right)^2
\left(\begin{array}{ccc}
l_f&2&l_i \\
-m_f&-2&m_i
\end{array}\right)^{-2}R_{i\rightarrow f}\, ,
\label{ml+2}
\end{equation}
where $R_{i\rightarrow f}$ corresponds to the case $\Delta l = 0$ and $\Delta m_l = -2$. 

\subsection{$\Delta l= +2$}

The only case we are left to consider is $\Delta l= +2$. If, for now, we only allow the transitions where $\Delta m_l= +2$, then all the calculations, except for the angular part, proceed in a manner very similar to the first case considered in this Appendix. However,  the angular integrals are straightforward to evaluate. Therefore we just present the result of the calculation. The  transition rate in the case $\Delta n= -k$, $\Delta l= +2$ and $\Delta m_l= +2$ for large $n$ is:
\begin{eqnarray} 
\label{Rif4}
R_{i\rightarrow f} &=&-  \frac{1}{24}\left(\frac{1}{n^2}-\frac{1}{(n-k)^2}\right) \frac{(2l+2)(2l+4)(n+l)!(n-k+l+2)!}{(2l+3)(2l+5) n^{2l+4}(n-k)^{2l+8}} \times \nonumber \\ 
& &~(n-l-1)!(n-k-l-3)! \Biggl\lbrace\sum_{m=0}^{n-l-1}\frac{(-1)^m}{(n-l-m-1)!(2l+m+1)!m!n^m}  \times  \nonumber \\ 
& &~\sum_{m'=0}^{n-k-l-3}\frac{(-1)^{m'}}{(n-k-l-m'-3)!(2l+m'+5)!m'!(n-k)^{m'}}\left[ \frac{2n(n-k)}{2n-k}^{2l+m+m'+4}\right]  \times \nonumber \\
& &~\Biggl[\frac{ (2l+m+m'+4)!}{4n-k} - \left( \frac{l+m'+2}{2n} + \frac{l+m}{2n-2k}+\frac{1}{2}\right)(2l+m+m'+3)!+ \nonumber \\ 
& &~ \frac{2n-k}{2n(n-k)} \left[ (l+m'+2)(l+m) - \frac{1}{2} \right] (2l+m+m'+2)! \Biggl]\Biggl\rbrace^2\, .
\end{eqnarray}

As an example, let us now consider a transition between two neighbouring states from a state $\vert i\rangle = \vert 100, 50, 50\rangle$ to the state $\vert i\rangle = \vert 99, 52, 52\rangle$. We have: 
\begin{equation} 
\label{numRif}
R_{i\rightarrow f} \approx 8.30\times10^{23}~\frac{1}{\mathrm{s}} 
\end{equation}
and the corresponding life time is 
\begin{equation} 
\label{numtau}
\tau \approx 1.20\times10^{-24}~\mathrm{s}\, . 
\end{equation}

When $\Delta m_l \neq +2$, we may calculate transition rates from Eq.~(\ref{WE}). If $\Delta m_l=-2$ one can easily show that the transition rate is given by 
\begin{equation}
\tilde R_{i\rightarrow f} = \left(\begin{array}{ccc}
l_f&2&l_i \\
-\tilde m_f&-2&\tilde m_i
\end{array}\right)^2
\left(\begin{array}{ccc}
l_f&2&l_i \\
-m_f&2&m_i
\end{array}\right)^{-2}R_{i\rightarrow f}\, ,
\label{ml+2}
\end{equation}
and in the same way one finds that  the case $\Delta m_l =0$ is related to the case $\Delta l=+2$,  and $\Delta m_l =-2$ in the following manner:
\begin{equation}
\tilde R_{i\rightarrow f} = \left(\begin{array}{ccc}
l_f&2&l_i \\
-\tilde m_f&0&\tilde m_i
\end{array}\right)^2
\left(\begin{array}{ccc}
l_f&2&l_i \\
-m_f&2&m_i
\end{array}\right)^{-2}R_{i\rightarrow f}\, .
\label{ml+2}
\end{equation}
These relations conclude our considerations in these particular cases.

To answer to the problem concerning the discreteness of the energy spectrum of the microscopic  black hole pairs, we have calculated the life time $\tau$ of the initial state of the system when all the possible transitions from a fixed initial state are taken into account. We have chosen the initial state to be  $\vert i\rangle = \vert 100, 50, 50\rangle$. This initial state yields a summation of about three hundred transition rates, when all the allowed G1 transitions are taken into account. A numerical estimate for each transition rate can be calculated by using the expressions for transition rates given in this Appendix. The life time of the initial state is obtained as an inverse of the sum of all the possible transition rates. We have found that when all the transitions are taken into account, the life time of the initial state is
\begin{equation}
\tau \approx 1.14 \times 10^{-30} {\mathrm s} \, .
\label{life}
\end{equation}

 \begin{figure}
    \begin{center}
      \leavevmode
      \epsfysize=55mm
      \epsfbox{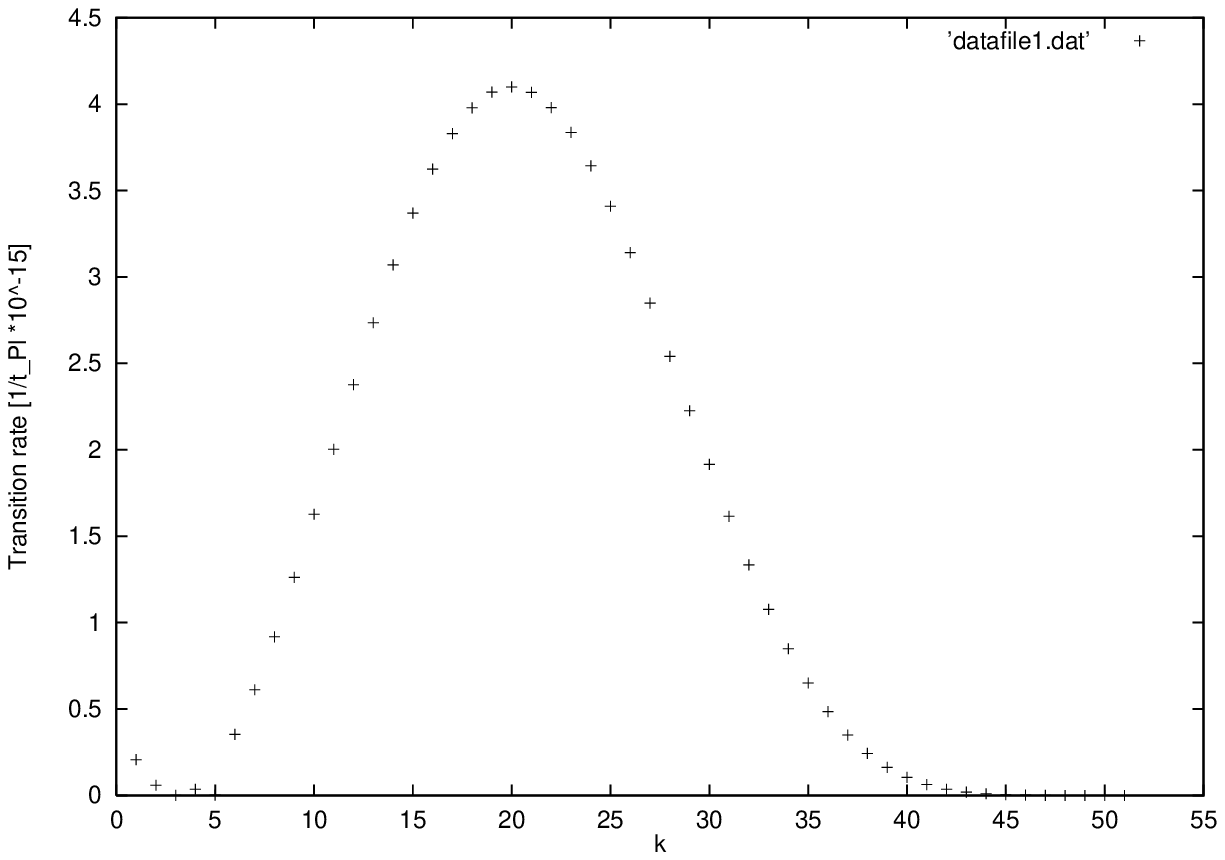}
    \end{center}
  \caption{The integrated transition rate $R_{i\rightarrow f}$ as a function of $k$, which is the difference between the initial value $n_i$ and the final value $n_f$ of the principal quantum number $n$. In this figure we have $\Delta l=\Delta m_l =-2$. The initial values of $n$, $l$ and $m_l$ are, respectively, $n_i=100$, $l_i=50$ and $m_{l_i}=50$. In all of the figures $R_{i\rightarrow f}$ has been plotted in the units of $\frac{1}{2t_{\mathrm{Pl}}} \times 10^{-15}$, where $t_{\mathrm{Pl}} := \sqrt{\frac{\hbar G}{c^5}} \approx 5.4 \times 10^{-44}~\mathrm{s}$ is the Planck time.}
  \label{fig1}
 \end{figure}

 \begin{figure}
    \begin{center}
      \leavevmode
      \epsfysize=55mm
      \epsfbox{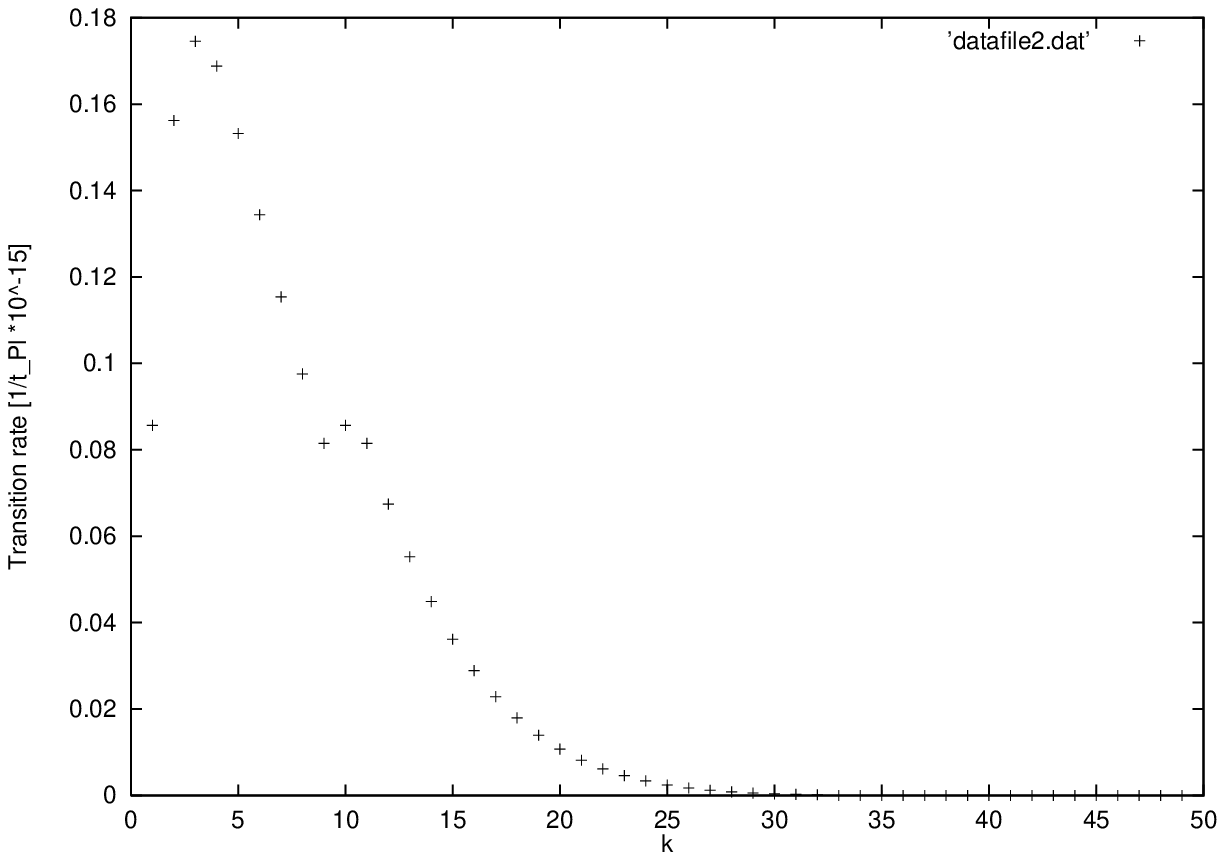}
    \end{center}
  \caption{$R_{i\rightarrow f}$ as a function of $k$, when $n_i=100$, $l_i=m_{l_i}=50$, $\Delta l=+2$, and $\Delta m_l =0$. }
  \label{fig2}
 \end{figure}

 \begin{figure}
    \begin{center}
      \leavevmode
      \epsfysize=55mm
      \epsfbox{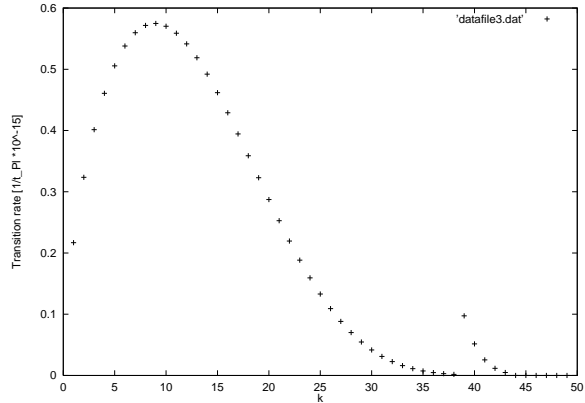}
    \end{center}
  \caption{$R_{i\rightarrow f}$ as a function of $k$, when $n_i=100$, $l_i=m_{l_i}=50$, $\Delta l=0$, and $\Delta m_l =-2$. }
  \label{fig3}
 \end{figure}


\begin{references}

\bibitem{Haw1} 
S.~W. Hawking, G.~T. Horowitz and S.~F. Ross, Phys.\ Rev.\ {\bf D51}, 4032 (1995).

\bibitem{Haw2}
S.~W. Hawking and S.~F. Ross, Phys.\ Rev.\ Lett.\ {\bf 75}, 3382 (1995).

\bibitem{Haw3}
S.~W. Hawking, Phys.\ Rev.\ {\bf D53}, 3099 (1996).

\bibitem{Haw4}
S.~W. Hawking and S.~F. Ross, Phys.\ Rev.\ {\bf D56}, 6403 (1997)

\bibitem{Haw5}
S.~W. Hawking in: {\it The Nature of Space and Time}, ed. by S.~W. Hawking
and R. Penrose (Princeton University Press, Princeton, 1996).

\bibitem{Messiah}
See, for example, A. Messiah: {\it Quantum Mechanics, Vol. 2}(North-Holland Physics Publishing Company, Amsterdam, 1986). 

\bibitem{Wald}
See, for example, R.~M. Wald: {\it Quantum Field Theory in Curved
Spacetime and Black Hole Thermodynamics} (The University of Chicago Press,
Chicago 1994).

\bibitem{Louko}
J. Louko and J. M\"{a}kel\"{a}, Phys.\ Rev.\ {\bf D54}, 4982 (1996).

\bibitem{Makela}
J. M\"{a}kel\"{a} and P. Repo, Phys.\ Rev.\ {\bf D57}, 4899 (1998).

\bibitem{Feynman}
A comprehensive account of quantization of linearized gravity is given in: {\it Feynman Lectures on Gravitation} by F.~B. Morinigo, W.~G. Wagner and R.~P. Feynman (Perseus Press, 1995).

\bibitem{Arfken}
See, for instance, G.~F. Arfken: {\it Mathematical Methods for Physicists} (Academic Press,
New York 1970).

\end{references}
\end{document}